\documentclass[conference]{IEEEtran}
\IEEEoverridecommandlockouts
\usepackage{cite}
\usepackage{amsmath,amssymb,amsfonts}
\usepackage{algorithm}
\usepackage{algpseudocode}
\usepackage{graphicx}
\usepackage{textcomp}
\usepackage{xcolor}
\usepackage{import}
\usepackage{tikz}
\usepackage{pgfplots} 
\usepackage{optidef}
\usepackage{hyperref}
\pgfplotsset{compat=1.17} 
\DeclareMathOperator*{\argmax}{argmax} 

\def\BibTeX{{\rm B\kern-.05em{\sc i\kern-.025em b}\kern-.08em
    T\kern-.1667em\lower.7ex\hbox{E}\kern-.125emX}}
\begin{document}

\title{Enhancing Sensing Privacy in ISAC Through Joint Signal and Artificial Noise Beamforming\\

\thanks{This work was supported by the National Science Foundation under grants 2348826, 2343465, 2343469 and 2418106.}
}

\author{\IEEEauthorblockN{Ahmad Musallam}
\IEEEauthorblockA{\textit{School of Aeronautics and Astronautics} \\
\textit{Purdue University}\\
West Lafayette, IN, USA \\
amusalla@purdue.edu}
\and
\IEEEauthorblockN{Husheng Li}
\IEEEauthorblockA{\textit{School of Aeronautics and Astronautics} \\
\textit{Purdue University}\\
West Lafayette, IN, USA \\
husheng@purdue.edu}
}

\maketitle

\begin{abstract}
Integrated sensing and communications (ISAC) is a promising feature in 6G networks. It is envisioned to enhance spectral efficiency and provide sensing and communication services that meet the stringent requirements of future applications. However, it also poses new security and privacy concerns by giving malicious attackers access to new information about the network. In this work, we focus on the sensing privacy of a monostatic ISAC system by investigating the capability of a sensing eavesdropper (EVE) with an unknown location, acting as a passive bistatic radar (PBR) to gain access to user location information. We then propose a joint transmit and artificial noise (AN) beamforming optimization problem to degrade EVE's performance. Finally, we propose an iterative algorithm to solve the proposed optimization problem and evaluate its performance.
\end{abstract}


\section{Introduction}

The limited availability of radio spectrum has increasingly become a challenge in recent years \cite{6651949}, as communication and radar functions require a greater bandwidth allocation to meet data rate and sensing accuracy requirements for various applications, such as autonomous driving, smart factories, and UAV navigation in GPS-denied environments \cite{9606831}. To address this challenge, significant research efforts have focused on innovative solutions. One promising paradigm is integrated sensing and communications (ISAC), which enables sensing and communication functionalities to share resources at various levels of integration, ranging from sharing common hardware to full integration under a common waveform. The latter is achieved by utilizing waveforms that meet both the sensing and communication requirements of the system \cite{10147248,8828023}, thus improving the spectrum and power efficiencies. 

However, ISAC’s design requirements introduce new security and privacy challenges for both communication and sensing functions ~\cite{10418473,10574259}. On the communication side, existing security threats are compounded by the risk of information leakage from the sensing link. Meanwhile, the favorable sensing characteristics of ISAC signals enable eavesdroppers (EVEs) to estimate channel state information (CSI) more accurately, granting them unauthorized access to sensitive user data, such as positions, movements, and actions. Addressing these combined challenges is essential for secure ISAC deployment in future networks.  

Recent works have focused on utilizing the accurate CSI provided by the ISAC system to improve communication security by employing physical layer security (PLS) measures \cite{10375133,10289830,10639496,10293761,10193088}. On the other hand, sensing privacy remains underexplored, despite being a major concern that needs to be addressed. Not only does it expose private user location information, but it also opens new avenues for active attacks against the ISAC network by allowing the malicious actor to gain better estimates of the CSI through access to the network geometry and better knowledge of the surrounding environment. In \cite{10465106} the authors evaluated the sensing capability of an internal adversary in a multi-static cell-free ISAC network and then proposed a transmit precoder design that jointly optimizes the sensing and communication requirements. Since sensing is a physical phenomenon, the most natural approach to ensuring sensing privacy would be through PLS measures. For example, the authors in \cite{10587082} considered a bistatic ISAC system where a single-antenna communication UE acts as a sensing EVE. They then proposed a mutual information (MI)-based optimization problem in order to maximize the legitimate receiver's MI and minimize the MI at EVE. Furthermore, they considered artificial noise-aided beamforming to further reduce EVE's performance based on knowledge of EVE's location and its CSI. In 
\cite{2408.11398}, the authors proposed a method based on generative artificial intelligence (GAI) to generate safeguarding signals that are modulated into the pilots at the transmitter in order to mask human activity from unauthorized listeners in an indoor environment. Jointly designing communication and sensing privacy and security measures is an approach that fits within the ISAC paradigm because one needs to consider both the sensing performance of the system and the communication rate at the UE. To this end, the authors in \cite{10605793} formulated an optimization problem where they tried to maximize the probability of target detection subject to a minimum signal-to-interference-plus-noise ratio (SINR) at the UE in the presence of a sensing and communications EVE.

In this work, we consider a monostatic ISAC system where an ISAC base station (BS) simultaneously provides communication and sensing services to the UE, with the presence of an EVE with unknown location that is trying to gain illegitimate access to the UE's position and actions. Specifically, we make the following contributions:
\begin{itemize}
    \item Investigate the ability of an EVE acting as a passive bistatic radar (PBR) to gain access to the UE position through interception of the signal reflected by the UE and a reference signal from the direct link between the BS and the EVE.
    \item formulate a joint signal and artificial noise beamforming optimization problem under a fixed power budget and ISAC system performance requirements to degrade the capability of a sensing EVE with an unknown location.
    \item propose an iterative algorithm to solve the proposed optimization problem based on the Dinkelbach method and successive convex approximation (SCA) and then showcase the performance of the proposed algorithm through numerical results.
\end{itemize}

The remainder of this paper is organized as follows. Section II describes the ISAC system model, the EVE sensing model, and the proposed problem formulation. In Section III, we formulate the optimization problem and its solution algorithm. Section IV presents the numerical results and discussion, and Section V concludes the paper.

\section{System Model and Problem Formulation}
\begin{figure}
    \centering
    \includegraphics[width=0.8\linewidth]{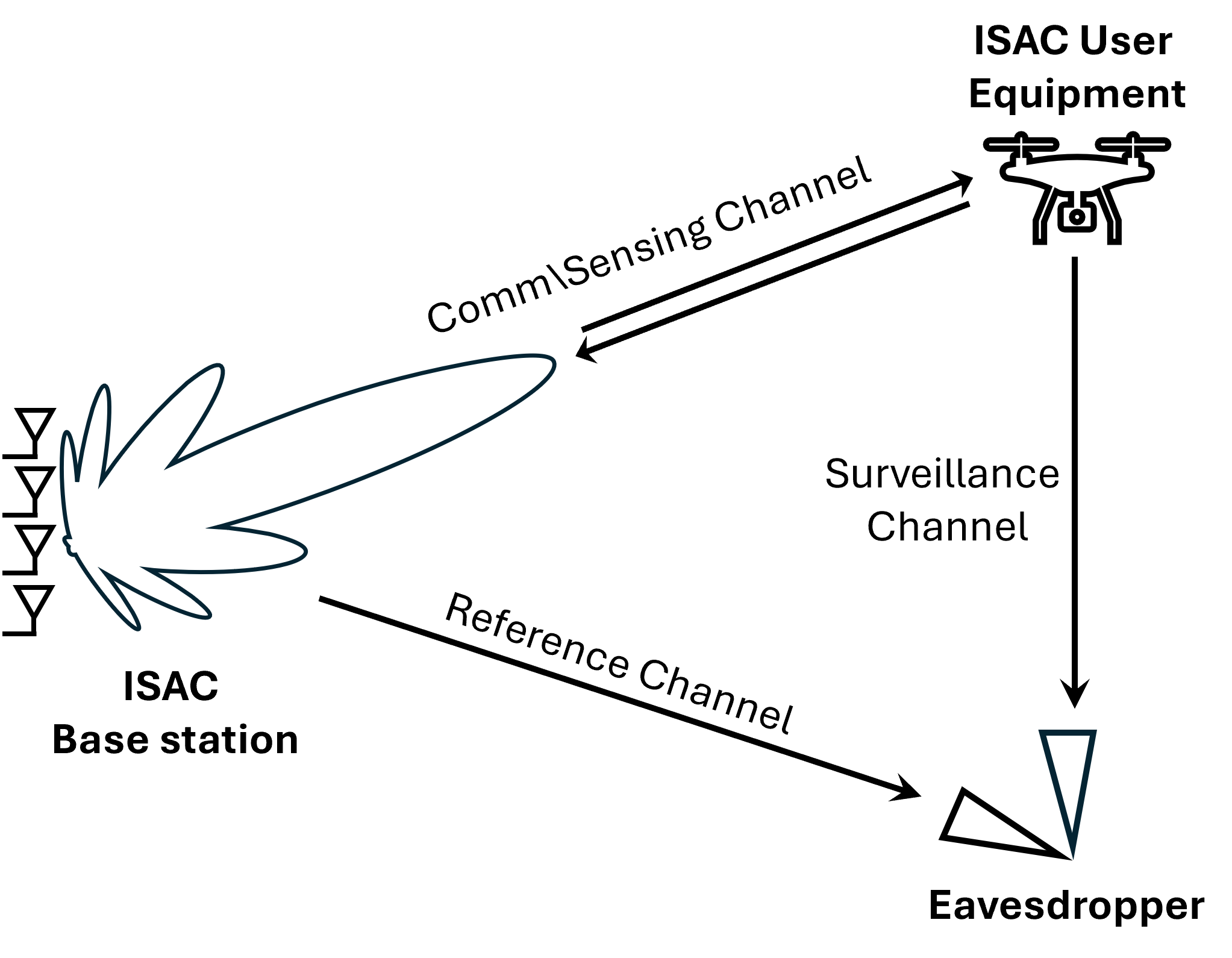}
    \caption{System model of ISAC and EVE}
    \label{fig:enter-label}
\end{figure}

We consider an ISAC BS, which simultaneously transmits a communication signal to a single-antenna UE and utilizes the reflected signal for monostatic radar sensing, with the goal of providing positioning services to the same UE. An unauthorized EVE, acting as a PBR, attempts to exploit the BS's transmitted signal to determine the location of the UE. EVE employs two sets of antennas: one directed towards the BS to capture a reference signal and the other aimed at the area of interest to intercept signals reflected from surrounding targets, as shown in Fig. \ref{fig:enter-label}.

In this scenario, we assume that the EVE is aware of its own position and that of the BS, while the BS does not have knowledge of EVE's location. We also assume that the BS knows the direction of the UE, which can be obtained through uplink communication.

\subsection{Signal Model}
We assume that the BS is equipped with a uniform linear array (ULA) with $N_t$ transmit antennas that transmit a common orthogonal frequency division multiplexing (OFDM) signal with $N$ subcarriers and $M$ data blocks per OFDM frame, and that the duration of each transmission block is $T_{sym} = T_{cp} + T$, where $T$ and $T_{cp}$,  respectively, denote the duration of the base OFDM symbol and the cyclic prefix (CP). The subcarrier spacing is $\Delta f = 1/T$, and the total bandwidth is $B=N\Delta f$. The transmit signal from the $i^{th}$ antenna element can be modeled as \cite{5393298}
\begin{equation}
    \mathbf{x}_{i}(t) = \sum_{m=0}^{M-1} \sum_{n=0}^{N-1}  w_i s_{n,m}e^{j2\pi n \Delta f (t-mT_{sym})}q(t-mT_{sym}), 
\end{equation}
where $s_{k,m}$ is the data symbol on the $n^{th}$ subcarrier and $m^{th}$ OFDM block, $\mathbf{w} = [w_1,w_2,\cdots, w_{Nt}]^T$ is the transmit beamforming vector, and $q(t)$ is the rectangular window that can be expressed as
\begin{equation}
    q(t) = \begin{cases}
    1, & t\in [-T_{cp},T]\\
    0, & \text{otherwise}
    \end{cases}.
\end{equation}
The signal reflected by the target and received at EVE is given by 
\begin{equation}
    y_{surv}(t) = \sum_{i=0}^{N_t-1} \alpha_{surv} a_i^*(\theta_u){x_i}(t-\tau_{surv})e^{j2\pi f_D t} + z_{surv}(t)
\end{equation}
where $\alpha_{surv}$, $\tau_{surv}$, and $f_D$ are the amplitude, delay, and Doppler frequency of the target signal, respectively. $z_{surv}(t)$ is the complex additive white Gaussian noise following $\mathcal{CN}(0,\sigma^2)$, and $\mathbf{a}(\theta)$ is the ULA steering vector, which is calculated at $\theta_u$ (the direction of the UE relative to the BS) and is given by
\begin{equation}
    \textbf{a}(\theta) \triangleq [1,e^{-j2\pi \frac{d}{\lambda}\sin{\theta}}, \hdots,e^{-j2\pi (N_{rx}-1) \frac{d}{\lambda}\sin{\theta}}]^T,
\end{equation}
where $d$ is the spacing of the ULA elements, and $\lambda = c_0/f_c$ is the carrier wavelength, where $c_0$ is the speed of light and $f_c$ is the carrier frequency. For the surveillance signal, $\alpha_{surv}$ is the bistatic radar gain, which is defined as
\begin{equation}
     \alpha_{surv} = \sqrt{\frac{\lambda^2 \sigma_{RCS}}{(4\pi)^3R_1^2R_2^2}},
\end{equation}
where $\sigma_{RCS}$ is the radar cross-section (RCS) of the UE, $R_1$ is the distance between the BS and the UE, and $R_2$ is the distance between the UE and EVE. 

Assuming that the BS and EVE are both stationary, the reference signal can be given by
\begin{equation}
    y_{ref}(t) = \sum_{i=0}^{N_t-1} \alpha_{ref} a_i^*(\theta_e){x_i}(t-\tau_{ref})+ z_{ref}(t),
\end{equation}
where $\theta_e$ is the angle between the BS and EVE,  $\tau_{ref}$ is the reference signal delay, $z_{ref}(t)$ is the complex additive white Gaussian noise following $\mathcal{CN}(0,\sigma^2)$, and $\alpha_{ref}$ is the amplitude of the reference channel due to free path loss over the distance between the BS and EVE ($R_{ref}$) and is given by
\begin{equation}
     \alpha = \sqrt{\lambda^2/(4\pi R_{ref}^2)},
\end{equation}

\subsection{Signal Processing at the Eavesdropper}

Since EVE is acting as passive bistatic radar, Target detection is achieved by calculating the cross-correlation between reference and surveillance signals, resulting in the cross-ambiguity function (CAF) defined as \cite{9098788}
\begin{equation}
    \beta(\tau_{rel},f_D) = \int_{-T_{tot}/2}^{T_{tot}/2} y_{ref}(t) \cdot y_{surv}(t-\tau_{rel}) \cdot e^{-j2\pi f_D t}dt,
\end{equation}
where $T_{tot}$ is the integration time, and $\tau_{rel} = \tau_{surv}-\tau_{ref}$ is the relative delay between the reference and surveillance signals. Since the true target range and delay are unknown, the CAF is calculated for a span of bistatic delays and Doppler frequencies, where the target is expected. The signals are coherently integrated in the CAF, leading to an integration gain when calculating the CAF. The signal-to-noise ratio (SNR) at the output of the CAF is given by
\begin{equation}
    \text{SNR}_{out} = \text{SNR}_{in}BT_{tot}, 
\end{equation}
which means that the integration gain is a function of the OFDM signal bandwidth $B=N\Delta f$ and the integration time $T_{tot} = MT_s$. For a fixed bandwidth and integration time, the output SNR, which determines the capability of EVE to detect targets, is governed by the SNR of the reference and surveillance signals. 

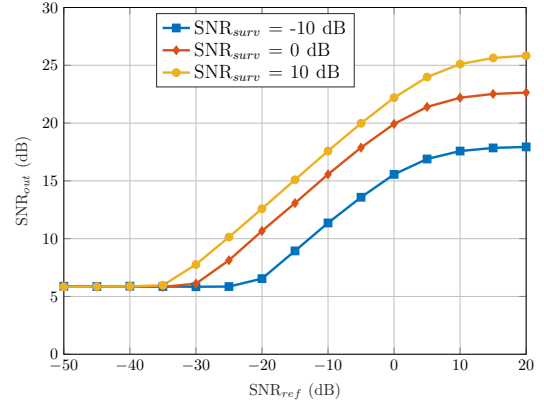
\begin{figure}[ht]
    \centering
    \resizebox{0.8\columnwidth}{!}{%
%
%
\definecolor{mycolor1}{rgb}{0.00000,0.44700,0.74100}%
\definecolor{mycolor2}{rgb}{0.85000,0.32500,0.09800}%
\definecolor{mycolor3}{rgb}{0.92900,0.69400,0.12500}%
\begin{tikzpicture}

\begin{axis}[%
width=4in,
height=3in,
at={(0.758in,0.481in)},
scale only axis,
xmin=-50,
xmax=20,
xlabel style={font=\normalsize\color{white!15!black}},
xlabel={$\text{SNR}_{ref}$ (dB)},
ymin=0,
ymax=30,
yminorticks=true,
ylabel style={font=\normalsize\color{white!15!black}},
ylabel={$\text{SNR}_{out}$ (dB)},
axis background/.style={fill=white},
xmajorgrids,
ymajorgrids,
legend style={font = \large, at={(0.2,0.77)}, anchor=south west, legend cell align=left, align=left, draw=white!15!black},
]

\addplot [color=mycolor1, line width=1.5pt, mark=square*, mark options={solid, fill=mycolor1, mycolor1}]
  table[row sep=crcr]{%
-50	5.88669294191519\\
-45	5.86236709099949\\
-40	5.88864743354442\\
-35	5.85767571486452\\
-30	5.84119687158228\\
-25	5.85588751536715\\
-20	6.54241084488385\\
-15	8.94134397779938\\
-10	11.3603002555972\\
-5	13.582392930608\\
0	15.5639170646378\\
5	16.8944961304083\\
10	17.5808467298697\\
15	17.8514330890233\\
20	17.9389793558005\\
};
\addlegendentry{$\text{SNR}_{surv}$ = -10 dB}

\addplot [color=mycolor2, line width=1.5pt, mark=diamond*, mark options={solid, fill=mycolor2, mycolor2}]
  table[row sep=crcr]{%
-50	5.86628254507808\\
-45	5.8503180618182\\
-40	5.8611336837555\\
-35	5.82633919747074\\
-30	6.10497639625665\\
-25	8.1305163823492\\
-20	10.6656893634494\\
-15	13.0778190154967\\
-10	15.5639414623759\\
-5	17.8816570208457\\
0	19.9243161690331\\
5	21.3987180155693\\
10	22.1937717507387\\
15	22.5216103734296\\
20	22.6365634398453\\
};
\addlegendentry{$\text{SNR}_{surv}$ = 0 dB}

\addplot [color=mycolor3, line width=1.5pt, mark=*, mark options={solid, fill=mycolor3, mycolor3}]
  table[row sep=crcr]{%
-50	5.84374160879097\\
-45	5.85313905410562\\
-40	5.87834681353632\\
-35	5.96955312230442\\
-30	7.76244445236867\\
-25	10.1344441477854\\
-20	12.5947579353578\\
-15	15.095607732698\\
-10	17.5838539284177\\
-5	19.9796823690957\\
0	22.1947419122986\\
5	23.9810383945107\\
10	25.1054136602882\\
15	25.6323435056534\\
20	25.8261113273111\\
};
\addlegendentry{$\text{SNR}_{surv}$ = 10 dB}

\end{axis}

\end{tikzpicture}%
    }
    \caption{Output SNR against $\text{SINR}_{ref}$ for different values of $\text{SNR}_{surv}$}
    \label{fig:pnfr}
\end{figure}

Fig. \ref{fig:pnfr} shows the effect of the signal-to-noise ratio in the reference channel ($\text{SNR}_{ref}$) and the surveillance channel ($\text{SNR}_{surv}$) on the output SNR of the CAF. It can be seen that as the SNR of the input signals decreases, the peak, which corresponds to the target, becomes lower until it reaches a point where the peak no longer corresponds to a target but rather to random noise. This is why the value becomes constant at around 6 dB.

For simplicity and without loss of generality, we are only considering a single target without clutter in this work. Hence, the maximum likelihood estimator for the target position corresponds to the CAF peak \cite{1455624}
\begin{equation*}
(\tau_{rel}^{target},f_D^{target}) = \argmax_{(\tau_{rel},f_D)}|\beta(\tau_{rel},f_D)|
\end{equation*}

\subsection{Artificial Noise Design}
In general, AN is designed to be randomly distributed to ensure that it is not correlated with the data signal. More specifically, it is designed to be an i.i.d. complex Gaussian random variable with zero mean and unit variance $\mathcal{CN}(0, \mathbf{I})$. Another approach for radar privacy would be to design the AN to either spoof the real target location or jam parts of the CAF. 

To spoof the target location by creating multiple false targets, the transmit noise signal needs to be comprised of the sum of delayed and Doppler shifted versions of the data signal, which can be modeled as
\begin{equation}
    \eta(t) = \sum_{k=1}^{K} \xi_kx(t-\tau_k) e^{j2\pi f_D^k t},
\end{equation}
where $\xi_k$, $\tau_k$, and $f_D^k$ are the amplitude, delay, and Doppler shift of the $k^{th}$ false target. 

The other approach would be to mask a region in the CAF by creating a peak at each delay and Doppler bin of the region of interest in the CAF according to the following.
\begin{equation}
    \eta(t) = \sum_{q=1}^{Q}\sum_{k=1}^{K} x(t-\tau_k) e^{j2\pi f_D^q t},
\end{equation}
where $k$ and $q$ are the indices of the $k^{th}$ delay and $q^{th}$ Doppler bin.

Both approaches can be combined with uncorrelated artificial noise to further reduce the chance of target detection by EVE.

\subsection{Problem Formulation}
Since our goal is to prevent EVE from sensing the location of the UE, we need to employ physical layer measures to reduce the SNR of the reference and surveillance channels. However, since the location of the EVE is unknown, techniques such as zero-forcing to suppress the reference signal are infeasible. In addition, introducing artificial noise (AN) into the surveillance channel will degrade the performance of the ISAC system. Thus, our proposed approach is to introduce artificial noise into the reference channel while simultaneously reducing the power of the data signal in all possible EVE directions. In order to achieve this, we will modify our transmit signal to be the superposition of beamformed data symbols and artificial noise. The transmit signal on the $i^{th}$ antenna is now given by 
\begin{equation}
    \mathbf{x}_{i}(t) = \sum_{m=0}^{M-1} \sum_{n=0}^{N-1}  \Tilde{s}_{n,m}^ie^{j2\pi n \Delta f (t-mT_{sym})}q(t-mT_{sym}), 
\end{equation}
where $\Tilde{s}_{n,m}^i$ is the transmit symbol from the $i^{th}$ antenna on the $n^{th}$ subcarrier and $m^{th}$ OFDM block, given by
\begin{equation}
    \Tilde{s}_{n,m}^i = w_i s_{n,m} + v_i \eta_{n,m},
\end{equation}
where $s_{n,m}$ and $\eta_{n,m}$ are the data and artificial noise symbols, $\mathbf{w} = [w_1,w_2,\cdots, w_{Nt}]^T$ is the data beamforming vector, and $\mathbf{v} = [v_1,v_2,\cdots, v_{Nt}]^T$ is the artificial noise beamforming vector.

\section{Joint Beamforming Optimization}
In this section, we investigate the problem of jointly optimizing the signal and artificial noise beamforming vectors, $\mathbf{w}$ and $\mathbf{v}$, under the assumption that the direction of EVE is unknown at the BS. To this end, we need to consider the worst-case scenario across all possible directions for EVE. We assume that the signal and artificial noise have unit power and that the transmit power is controlled by the beamforming vectors. This turns our problem into minimizing the maximum signal-to-artificial noise power ratio for all $\theta_e$. Our optimization problem can be formulated as
\begin{mini!}|s|
  {\mathbf{v}, \mathbf{w}}{\max_{\theta_e} \frac{|\mathbf{a}^H(\theta_e) \mathbf{w}|^2} {|\mathbf{a}^H(\theta_e) \mathbf{v}|^2}}{}{}
  \addConstraint{\|\mathbf{w}\|^2 + \|\mathbf{v}\|^2}{\leq P_{\text{total}}}\label{p1c1}
  \addConstraint{|\mathbf{a}^H(\theta_u) \mathbf{w}|^2}{\geq P_{\text{user}}}\label{p1c2}
  \addConstraint{|\mathbf{a}^H(\theta_u) \mathbf{v}|^2}{= 0}\label{p1c3}
\end{mini!}
where \eqref{p1c1} ensures the total transmit power constraint, while \eqref{p1c2} guarantees that the sensing and communication requirements for the user are met by specifying the transmit power in the direction of the user which can be either limited by the communication SNR at the UE or the sensing SNR at the BS, and \eqref{p1c3} ensures orthogonality between the UE direction and AN. 

Note that here we assumed that the noise figure (NF) for the EVE's receiver is unknown. However, this will not be an issue since the overall system performance will be better than predicted. This is because the noise will be a constant term added to the denominator of the objective function, which will cause the overall SINR at EVE to be lower. 

In general, it is difficult to directly solve problem (12) due to the non-convexity of the objective function and the constraint \eqref{p1c2} as well as the equality constraint in \eqref{p1c3}. Furthermore, the variable $\theta_e$ in the objective is continuous, making it more challenging to optimize. We start by addressing the objective function, where we discretize $\theta_e$ over the range of angles of potential eavesdropper directions. Next, we define a matrix $\mathbf{A} \triangleq [\overline{\mathbf{a}(\theta_{e,1})},\overline{\mathbf{a}(\theta_{e,2})},\cdots,\overline{\mathbf{a}(\theta_{e,L})}]^T$, where each row represents the conjugate of the steering vector $\mathbf{a}(\theta)$ at each discrete angle $\theta_{e,i}$. Thus, our objective function can now be defined as
\begin{equation}\label{f_v_w}
    f(\mathbf{v},\mathbf{w}) \triangleq \max \biggl\{(\overline{\mathbf{Aw}} \odot \mathbf{Aw}) \oslash (\overline{\mathbf{Av}} \odot \mathbf{Av}) \biggr\},
\end{equation}
where $\odot$ and $\oslash$ are, respectively, element-wise vector multiplication and division. We can then use the Dinkelbach method \cite{Phillips2009} to turn the objective function into a difference of two convex functions as follows:  
\begin{equation}
g(\mathbf{v},\mathbf{w}, \lambda) = \max  \biggl\{(\overline{\mathbf{Aw}} \odot \mathbf{Aw}) - \lambda (\overline{\mathbf{Av}} \odot \mathbf{Av}) \biggr\},
\end{equation}
where the problem becomes an iterative optimization problem in which we try to find the optimal $\mathbf{w^*}$, $\mathbf{v^*}$, and $\lambda^*$ that satisfy 
\begin{equation}
g(\mathbf{v^*},\mathbf{w^*}, \lambda^*) = 0.
\end{equation}

However, the problem remains non-convex. To ensure convexity, we apply successive convex approximation (SCA) by approximating the second term with its first-order Taylor series expansion \cite{doi:10.1137/120891009}. This transforms our optimization subproblem into an iterative solution of a convex problem, where we solve a convex problem at each iteration. Our new objective function can now be written as
\begin{equation}
\begin{split}
\Tilde{g}(\mathbf{v},\mathbf{w}, \lambda) = \max \biggl\{(\overline{\mathbf{Aw}} \odot \mathbf{Aw}) - \lambda \biggl( \overline{\mathbf{A\Tilde{v}}} \odot \mathbf{A\Tilde{v}} + \\
2\operatorname{\mathbb{R}e}\left\{ \overline{\mathbf{A\Tilde{v}}} \odot \mathbf{A(v-\Tilde{v})} \right\} \biggr) \biggl\},
\end{split}
\end{equation}
where $\mathbf{\Tilde{w}}$ and $\mathbf{\Tilde{v}}$ are the values of $\mathbf{w}$ and $\mathbf{v}$ from the previous iteration. We can follow the same procedure to turn the constraint in \eqref{p1c2} into a convex constraint as follows:  
\begin{equation}\label{p2c2}
    |\mathbf{a}^H(\theta_u) \mathbf{\Tilde{w}}|^2 + 2\operatorname{\mathbb{R}e}\left\{\mathbf{a}^H(\theta_u) (\mathbf{w}-\mathbf{\Tilde{w}})\right\} \geq P_{\text{user}}.
\end{equation}

 Finally, we can relax constraint \eqref{p1c3} by ensuring that the right-hand side is less than or equal to a small constant $\epsilon$.
\begin{equation}\label{p2c3}
    |\mathbf{a}^H(\theta_u) \mathbf{v}|^2 \leq \epsilon.
\end{equation}

The optimization subproblem can now be expressed as:
\begin{mini!}|s|
    {\mathbf{w}, \mathbf{v}}{ \Tilde{g}({\mathbf{w}, \mathbf{v},\lambda})}{}{}
    \addConstraint{\eqref{p1c1}, \eqref{p2c2}, \eqref{p2c3}}{},
\end{mini!}
which can easily be shown to be convex, and can be used to iteratively solve for the optimal beamforming vectors $\mathbf{w}$ and $\mathbf{v}$. The iterative solution procedure for the problem in (21) is presented in Algorithm \ref{alg:alg1}.

\begin{algorithm}
    \caption{Iterative Algorithm for Solving Problem (21)}\label{alg:alg1}
    \begin{algorithmic}[1]
        \State initialize $\mathbf{w}^{(0)},\mathbf{v}^{(0)}, \lambda^{(0)}<1, i=0, \delta_1, \delta_2$ 
            \Repeat
                \State $i \gets i+1$
                \State initialize $j=0$
                \State $\hat{\mathbf{w}}^{(0)},\hat{\mathbf{v}}^{(0)} \gets \mathbf{w}^{(i-1)},\mathbf{v}^{(i-1)}$
                \Repeat
                \State $j \gets j+1$
                \State $\Tilde{\mathbf{w}}^{(j)},\Tilde{\mathbf{v}}^{(j)} \gets \hat{\mathbf{w}}^{(j-1)},\hat{\mathbf{v}}^{(j-1)}$
                \State Solve problem (21) to get the optimal 
                $\hat{\mathbf{w}}^{(j)},\hat{\mathbf{v}}^{(j)}$
                \Until{$|f(\hat{\mathbf{w}}^{(j)},\hat{\mathbf{v}}^{(j)}) - f(\hat{\mathbf{w}}^{(j-1)},\hat{\mathbf{v}}^{(j-1)})| < \delta_1$ }
                \State $\mathbf{w}^{(i)},\mathbf{v}^{(i)} \gets \hat{\mathbf{w}}^{(j)},\hat{\mathbf{v}}^{(j)}$
                \State $\lambda^{(i)} \gets f(\mathbf{w}^{(i)},\mathbf{v}^{(i)})$
            \Until{$g(\mathbf{w}^{(i)},\mathbf{v}^{(i)},\lambda^{(i)}) < \delta_2$ }\\
        \Return $\mathbf{w}^{(i)},\mathbf{v}^{(i)}$
    \end{algorithmic}
\end{algorithm}

We note that with each iteration of algorithm \ref{alg:alg1} it is guaranteed that the value of the objective function is less than or equal to the previous iteration (i.e. $g^{i+1} \leq g^{i}$); hence, the algorithm is guaranteed to converge to a local minimum. Furthermore, each inner loop iteration solves a convex optimization problem of size $\mathcal{O}(N_t^{3.5})$ and computes the function in \eqref{f_v_w} that has a size of $\mathcal{O}(LN_t)$, where L is the number of discrete values of $\mathbf{a(\theta_e)}$. Therefore, the total computational complexity of Algorithm \ref{alg:alg1} is $\mathcal{O}((N_t^{3.5}+LN_t)\cdot log(1/\delta_1)\cdot log(1/\delta_2))$, where $\delta_1$ and $\delta_2$ are, respectively, the tolerance in the outer and inner loops. Therefore, assuming that $L < N_t^{2.5}$, the final complexity of the algorithm is $\mathcal{O}(N_t^{3.5}\cdot log(1/\delta_1)\cdot log(1/\delta_2))$.

\section{Numerical Results}
In this section, we provide simulation results to evaluate the performance of the proposed algorithm in improving sensing privacy in a monostatic ISAC system under a fixed power budget and ISAC performance requirements. The parameters of the simulated ISAC system are shown in Table \ref{tab:param_table}.
\subsection{Random Artificial Noise}
\begin{table}[ht]
\begin{center}
    \caption{parameters for the simulated ISAC system}
    \begin{tabular} { c  c }
        \hline
        \textbf{Parameter} & \textbf{Value}\\
        \hline
        Bandwidth & 50 MHz \\
        Carrier Frequency ($f_c$) & 5.5 GHz\\
        number of OFDM subcarriers ($N$) & 1024\\
        number of OFDM blocks ($M$) & 64 \\
        UE direction ($\theta_u$) & $\in [-\pi/2,\pi/2]$\\ 
        EVE direction ($\theta_e$) & $\in [-\pi/2,\pi/2]/\theta_u$\\ 
        $P_{user}$ & 30 dBm\\
        \hline
    \end{tabular}
    \label{tab:param_table}
\end{center}
\end{table}

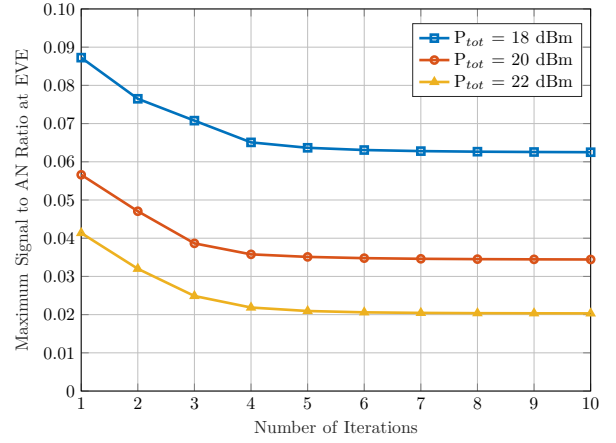
\begin{figure}
    \centering
    \resizebox{0.9\columnwidth}{!}{%
%
%
\definecolor{mycolor1}{rgb}{0.00000,0.44700,0.74100}%
\definecolor{mycolor2}{rgb}{0.85000,0.32500,0.09800}%
\definecolor{mycolor3}{rgb}{0.92900,0.69400,0.12500}%
\begin{tikzpicture}

\begin{axis}[%
width=4in,
height=3in,
at={(0.758in,0.481in)},
scale only axis,
xmin=1,
xmax=10,
xtick={ 1,  2,  3,  4,  5,  6,  7,  8,  9, 10, 11, 12, 13, 14, 15},
xlabel style={font=\normalsize\color{white!15!black}},
xlabel={Number of Iterations},
ymin=0.0,
ymax=0.1,
ytick = {0, 0.01, 0.02, 0.03, 0.04, 0.05, 0.06, 0.07, 0.08, 0.09, 0.10},
yticklabels = {0, 0.01, 0.02, 0.03, 0.04, 0.05, 0.06, 0.07, 0.08, 0.09, 0.10},
ylabel style={font=\normalsize\color{white!15!black}},
ylabel={Maximum Signal to AN Ratio at EVE},
axis background/.style={fill=white},
xmajorgrids,
ymajorgrids,
legend style={font = \normalsize, at={(0.65,0.77)}, anchor=south west, legend cell align=left, align=left, draw=white!15!black},
]

\addplot [color=mycolor1, line width=1.5pt, mark=square, mark options={solid, fill=mycolor1, mycolor1}]
  table[row sep=crcr]{%
1	0.0872552009863622\\
2	0.0764884355745645\\
3	0.0707702592587208\\
4	0.0650808288218001\\
5	0.0636617098486931\\
6	0.0630783854160077\\
7	0.0627965094406022\\
8	0.0626450555602445\\
9	0.0625619536120798\\
10	0.0625055302363199\\
};
\addlegendentry{$\text{P}_{tot}$ = 18 dBm}

\addplot [color=mycolor2, line width=1.5pt, mark=o, mark options={solid, fill=mycolor2, mycolor2}]
  table[row sep=crcr]{%
1	0.0566002778453104\\
2	0.0470572822062143\\
3	0.038651154503241\\
4	0.0357758793213719\\
5	0.0351094695223343\\
6	0.0347787974822899\\
7	0.0346113166905584\\
8	0.0345205583290549\\
9	0.0344663162072557\\
10	0.0344320395579613\\
};
\addlegendentry{$\text{P}_{tot}$ = 20 dBm}

\addplot [color=mycolor3, line width=1.5pt, mark=triangle, mark options={solid, fill=mycolor3, mycolor3}]
  table[row sep=crcr]{%
1	0.0413569463701986\\
2	0.0319969169148395\\
3	0.0248758789906876\\
4	0.0218559234316309\\
5	0.0209470606206183\\
6	0.0205973919384962\\
7	0.0204515636758305\\
8	0.0203760750064506\\
9	0.0203463209224681\\
10	0.0203227435322301\\
};
\addlegendentry{$\text{P}_{tot}$ = 22 dBm}

\end{axis}

\end{tikzpicture}%
    }
    \caption{Convergence of optimization algorithm indicated by change in objective function against number of iterations for different power budgets.}
    \label{fig:itr_vs_maxSANR}
\end{figure}

Fig. \ref{fig:itr_vs_maxSANR} shows the convergence performance of the proposed algorithm under different power budgets. It can be seen that all three cases converge in 6 or 7 iterations, which illustrates the fast convergence rate of the proposed algorithm. In addition, it shows that as the total transmit power increases, the objective function value decreases. This is because, for a fixed performance requirement of the ISAC system, there is more power available to allocate to artificial noise, which improves the objective function value.

\begin{figure}
    \centering
    \resizebox{0.9\columnwidth}{!}{%
%
%
\definecolor{mycolor3}{rgb}{0.00000,0.44700,0.74100}%
\definecolor{mycolor1}{rgb}{0.85000,0.32500,0.09800}%
\definecolor{mycolor2}{rgb}{0.92900,0.69400,0.12500}%
\begin{tikzpicture}

\begin{axis}[%
width=4in,
height=3in,
at={(0.684in,0.445in)},
scale only axis,
xmin=16,
xmax=25,
xlabel style={font=\normalsize\color{white!15!black}},
xlabel={Total Transmit Power (dBm)},
ymin=0,
ymax=1,
ylabel style={font=\normalsize\color{white!15!black}},
ylabel={$\text{Probability of Target Detection (P}_\text{d}\text{)}$ at EVE},
ytick={0,0.1, 0.2, 0.3, 0.4, 0.5, 0.6, 0.7, 0.8, 0.9, 1},
axis background/.style={fill=white},
xmajorgrids,
ymajorgrids,
legend style={font = \footnotesize, at={(0.0,0.0)}, anchor=south west, legend cell align=left, align=left, draw=white!15!black},
]

\addplot [color=mycolor3, line width=1.5pt, mark=triangle, mark options={solid, rotate=180, mycolor3}]
  table[row sep=crcr]{%
16	1\\
17	1\\
18	1\\
19	0.999\\
20	0.968\\
21	0.851\\
22	0.647\\
23	0.436\\
24	0.246\\
25	0.158\\
};
\addlegendentry{$\text{SNR}_\text{s}$ = 10 dB,$\text{SNR}_\text{r}$ = 10 dB}

\addplot [color=mycolor2, line width=1.5pt, mark=o, mark options={solid, mycolor2}]
  table[row sep=crcr]{%
16	1\\
17	1\\
18	0.999\\
19	0.986\\
20	0.915\\
21	0.736\\
22	0.485\\
23	0.317\\
24	0.180\\
25	0.107\\
};
\addlegendentry{$\text{SNR}_\text{s}$ = 10 dB,$\text{SNR}_\text{r}$ = 5 dB}

\addplot [color=mycolor1, line width=1.5pt, mark=square, mark options={solid, mycolor1}]
  table[row sep=crcr]{%
16	1\\
17	1\\
18	0.991\\
19	0.904\\
20	0.658\\
21	0.41\\
22	0.241\\
23	0.127\\
24	0.08\\
25	0.046\\
};
\addlegendentry{$\text{SNR}_\text{s}$ = 10 dB,$\text{SNR}_\text{r}$ = 0 dB}

\addplot [color=mycolor3, dashed, line width=1.5pt, mark=triangle, mark options={solid, rotate=180, mycolor3}]
  table[row sep=crcr]{%
16	1\\
17	1\\
18	0.992\\
19	0.891\\
20	0.671\\
21	0.395\\
22	0.225\\
23	0.146\\
24	0.084\\
25	0.044\\
};
\addlegendentry{$\text{SNR}_\text{s}$ = 0 dB,$\text{SNR}_\text{r}$ = 10 dB}

\addplot [color=mycolor2, dashed, line width=1.5pt, mark=o, mark options={solid, mycolor2}]
  table[row sep=crcr]{%
16	1\\
17	1\\
18	0.96\\
19	0.781\\
20	0.494\\
21	0.269\\
22	0.171\\
23	0.116\\
24	0.062\\
25	0.044\\
};
\addlegendentry{$\text{SNR}_\text{s}$ = 0 dB,$\text{SNR}_\text{r}$ = 5 dB}

\addplot [color=mycolor1, dashed, line width=1.5pt, mark=square, mark options={solid, mycolor1}]
  table[row sep=crcr]{%
16	1\\
17	0.988\\
18	0.755\\
19	0.44\\
20	0.216\\
21	0.149\\
22	0.091\\
23	0.054\\
24	0.019\\
25	0.02\\
};
\addlegendentry{$\text{SNR}_\text{s}$ = 0 dB,$\text{SNR}_\text{r}$ = 0 dB}

\end{axis}

\end{tikzpicture}%
    }
    \caption{Probability of target detection $P_d$ at EVE against total transmit power for different $\text{SNR}_e$ and $\text{SNR}_r$ values.}
    \label{fig:pwr_vs_pd}
\end{figure}
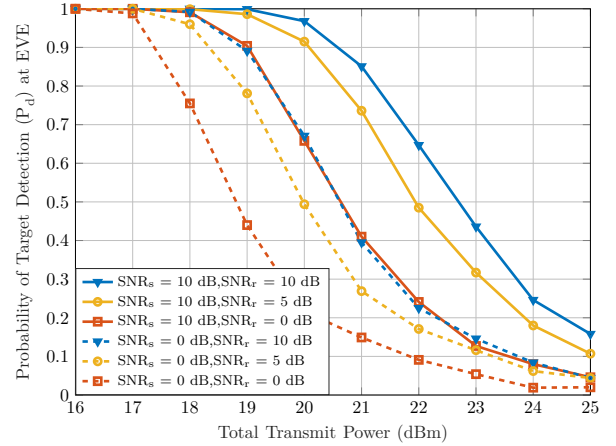

Fig. \ref{fig:pwr_vs_pd} demonstrates the capability of EVE in detecting the UE under different power budgets and SNR in the reference and surveillance channels for fixed performance requirements of the ISAC system. It can be seen that, for a low power budget, EVE is still able to detect the UE with a high percentage because the SNR at the output of the CAF is still higher than the detection threshold. However, as the power budget increases, the probability of detection decreases. Additionally, we can see the effect of the SNR in the reference and surveillance channels on EVE's target detection performance. It can be noted that an equal decrease in either $\text{SNR}_{ref}$ or $\text{SNR}_{surv}$ leads to the same degradation in EVE's performance.

Fig. \ref{fig:nt_vs_pd} shows the impact of the number of transmit antennas ($N_t$) on EVE's target detection probability. It can be seen that as the number of transmit antennas increases, the probability of detection of the target by EVE decreases. This is due to the increase in the degrees of freedom in the beamforming optimization, which allows the transmitter to produce a narrower main beam and further suppress the sidelobes in the data signal while, at the same time, having more control over the distribution of AN.

\subsection{Spoofing and Masking with AN}
We evaluated the performance of the various AN design approaches discussed in Section II-C, as illustrated in Fig. \ref{fig:spoof_jam}. In (a), the CAF without AN clearly shows the target position, while in (b) when a random AN is applied, the target peak becomes obscured. Figure (c) depicts a spoofing scenario in which AN was replaced with signals having artificial delays and Doppler shifts. We notice that multiple false targets have been detected by EVE. However, the true target position is still visible. Although the performance of random AN is better, we note that the total transmit power for this method is less; hence, if believable false targets are faked, it can make it difficult for EVE to distinguish true from false targets. Lastly, in (d), we tested masking ranges between 50 and 200 meters and relative velocities between -30 and 30 m/s. However, the actual mask was shifted to be centered around the target because the reference signal will be cross-correlated with the surveillance signal, which makes it possible to detect the target, despite the more challenges than in the case without AN. 
\begin{figure}
    \centering
    \resizebox{0.9\columnwidth}{!}{%
%
%
\definecolor{mycolor1}{rgb}{0.00000,0.44700,0.74100}%
\definecolor{mycolor2}{rgb}{0.85000,0.32500,0.09800}%
\definecolor{mycolor3}{rgb}{0.92900,0.69400,0.12500}%
\begin{tikzpicture}

\begin{axis}[%
width=4in,
height=3in,
at={(0.758in,0.509in)},
scale only axis,
xmin=16,
xmax=32,
xlabel style={font=\normalsize\color{white!15!black}},
xlabel={$\text{Number of Transmit Antenna (N}_\text{t}\text{)}$},
ymin=0,
ymax=0.35,
ylabel style={font=\normalsize\color{white!15!black}},
ylabel={$\text{Probability of Target Detection (P}_\text{d}\text{)}$ at EVE},
ytick = {0.00, 0.05, 0.10, 0.15, 0.20, 0.25, 0.30, 0.35},
yticklabels = {0, 0.05, 0.10, 0.15, 0.20, 0.25, 0.30, 0.35},
axis background/.style={fill=white},
xmajorgrids,
ymajorgrids,
legend style={legend cell align=left, align=left, draw=white!15!black}
]
\addplot [color=mycolor1, line width=1.5pt, mark=square, mark options={solid, mycolor1}]
  table[row sep=crcr]{%
16	0.3295\\
20	0.1274\\
24	0.0484\\
28	0.0114\\
32	0.0055\\
};
\addlegendentry{$\text{P}_{\text{tot}}\text{ = 32 dBm}$}

\addplot [color=mycolor2, line width=1.5pt, mark=o, mark options={solid, mycolor2}]
  table[row sep=crcr]{%
16	0.1833\\
20	0.0555\\
24	0.0223\\
28	0.0055\\
32	0.0044\\
};
\addlegendentry{$\text{P}_{\text{tot}}\text{ = 34 dBm}$}

\addplot [color=mycolor3, line width=1.5pt, mark=triangle, mark options={solid, rotate=180, mycolor3}]
  table[row sep=crcr]{%
16	0.0862\\
20	0.0227\\
24	0.0108\\
28	0.0047\\
32	0.0032\\
};
\addlegendentry{$\text{P}_{\text{tot}}\text{ = 36 dBm}$}

\end{axis}
\end{tikzpicture}%
    }
    \caption{Probability of target detection $P_d$ at EVE against number of transmit antenna ($N_t$) at different transmit power for $\text{SNR}_{ref} = \text{SNR}_{surv} = 10 \text{dB}$.}
    \label{fig:nt_vs_pd}
\end{figure}
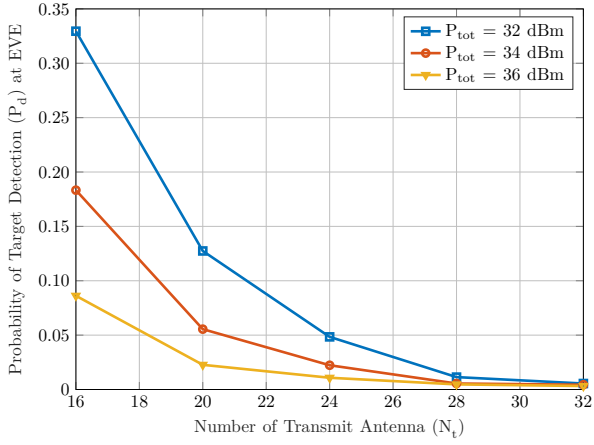
\begin{figure}
    \centering
    \resizebox{\columnwidth}{!}{%
     \includegraphics[]{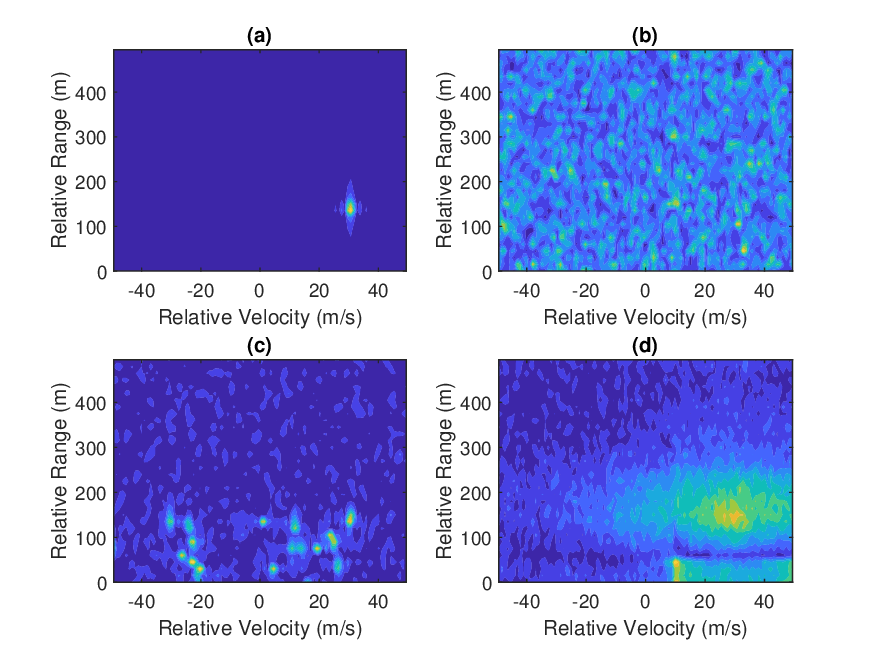}
     }
    \caption{The cross ambiguity function plot for the case (a) without AN (b) with random AN (c) spoofing (d) masking.}
    \label{fig:spoof_jam}
\end{figure}
\section{Conclusion}
In this work, we investigated the capability of a sensing EVE acting as a passive bistatic radar in a monostatic ISAC system. We then formulated an optimization problem to limit its access to UE location information by jointly optimizing the transmit and artificial noise beamforming vectors under a fixed power budget and ISAC system performance. Finally, we proposed an algorithm to solve the optimization problem and showed that the proposed algorithm converges to an optimal solution quickly and efficiently. The simulation results show that the proposed method is effective in reducing the EVE sensing capability.

\bibliographystyle{IEEEtran}
\bibliography{references}

\end{document}